\documentclass[preprint,tigten,letteredappendix,appendixfloats]{aastex631}

\usepackage{amsmath}
\usepackage{romannum}
\usepackage{natbib}
\usepackage{enumitem}
\usepackage{soul}
\usepackage{longtable}

\usepackage{graphicx}           
\usepackage{latexsym}           
\usepackage{bm}                 
\usepackage[english]{babel}
\usepackage{color}              

\def\arrvline{\hfil\kern\arraycolsep\vline\kern-\arraycolsep\hfilneg}

\newcommand{\acron}{RXDF}

\usepackage{fancyhdr}
\makeatletter
\let\ps@titlepage\ps@plain
\makeatother

\begin{document}

\pagenumbering{arabic}

\title{The Roman eXtreme Deep Field (RXDF)}

\author[0000-0001-7592-7714]{Haojing Yan}
\affiliation{Department of Physics and Astronomy, University of Missouri, Columbia, MO 65211, USA}
\email{yanha@missouri.edu}

\author[0000-0002-6610-2048]{Anton M. Koekemoer} 
\affiliation{Space Telescope Science Institute, Baltimore, MD 21218, USA}
\email{koekemoer@stsci.edu}

\author[0000-0003-1659-7035]{Yue Shen}
\affiliation{Department of Astronomy, University of Illinois Urbana-Champaign, Urbana, IL 61801, USA}
\email{shenyue@illinois.edu}

\author[0000-0001-7957-6202]{Bangzheng Sun}
\affiliation{Department of Physics and Astronomy, University of Missouri, Columbia, MO 65211, USA}

\author[0000-0001-9440-8872]{Norman A. Grogin} 
\affiliation{Space Telescope Science Institute, Baltimore, MD 21218, USA}

\author[0000-0002-8876-5248]{Zihao Wu}
\affiliation{Center for Astrophysics \textbar\ Harvard \& Smithsonian, Cambridge, MA 02138, USA}

\author[0000-0002-8896-6496]{Christian Kragh Jespersen}
\affiliation{Department of Astrophysical Sciences, Princeton University, Princeton, NJ 08544, USA}

\author[0000-0002-6748-6821]{Rachel Somerville}
\affiliation{Center for Computational Astrophysics, Flatiron Institute, New York, NY 10010, USA}

\author[0000-0003-3004-9596]{Kyoung-Soo Lee}
\affiliation{Department of Physics and Astronomy, Purdue University, West Lafayette, IN 47906, USA}

\author[0000-0002-8360-3880]{Dale D. Kocevski}
\affiliation{Department of Physics and Astronomy, Colby College, Waterville, ME 04901, USA}

% ...

\begin{abstract}

The Roman eXtreme Deep Field (\acron) program is one of the five General 
Astrophysics Survey (GAS) programs approved for observing time with the Nancy Grace 
Roman Space Telescope in Cycles 1 and 2. It has been allocated 386.41~hours to 
carry out an imaging survey to ${\rm AB} = 30$~mag (5$\sigma$) over 
$\sim$140$\times$ larger area than the Hubble eXtreme Deep Field (HXDF) 
full-depth area (ACS+WFC3/IR). The RXDF will cover the full Roman wavelength 
range with 7 bands, reaching ${\rm AB} = 30$~mag in $RZYJH$, 29~mag in $F$, and 
28~mag in $K$, over a full-depth area of 678.75~arcmin$^2$ embedded in a total 
area of 1,243~arcmin$^2$, and far exceeding the depths of the Roman Core 
Community Surveys (CCS). The RXDF is within the Euclid Ultra Deep Field (EUDF) 
near the North Ecliptic Pole (NEP), a strategic long-term field for generational 
space facilities, with a wealth of multi-wavelength data including extensive 
coverage from the James Webb Space Telescope (JWST) NEXUS Treasury program. The observations will cover 
3 epochs at a 1-year cadence, each epoch divided into 3 sub-epochs $\sim$10 days 
apart, enabling time-domain studies on time baselines from $\sim$10 days to over
$\sim$2 years. The RXDF is uniquely positioned to address critical questions in reionization, 
large scale structure (LSS), growth of supermassive black holes (SMBHs), little 
red dots (LRDs), and high-$z$ supernovae (SNe); the volumes probed  by HST+JWST 
are too small at these extreme depths, and even the deepest CCS tiers are too 
shallow. In addition to our key objectives, a wealth of additional science will 
be enabled by engaging the community with our rapidly released datasets, 
revolutionizing a wide range of science for a lasting legacy. This short 
document, which is converted from the approved \acron\ proposal, aims to provide 
the community with a summary of the program.

\end{abstract}


\section{Synopsis}

    JWST has made many new discoveries at the faintest magnitudes across 
all redshifts, but addressing the questions raised by these discoveries will 
require studies to AB $\sim$30~mag over a much larger area than JWST's reach. The Roman 
eXtreme Deep Field (\acron; Roman PID 2001, PI H. Yan, co-PIs A. Koekemoer and
Y. Shen) will answer this call by carrying out 7-band, 0.5--2.3~$\mu$m imaging in 
one WFI pointing in the North Ecliptic Pole (NEP) region within the Euclid
Ultra-Deep Field (EUDF), part of which is being observed by the North ecliptic 
pole EXtragalactic Unified Survey (NEXUS; \citealt{Shen2024nexus}), a JWST 
multicycle (Cycle 3--5) Treasury program (JWST PID 5105, PI Y. Shen).
\acron\, will reach 5$\sigma$ limits 
\footnote{Throughout this document, the depths are quoted as 5$\sigma$ limiting 
magnitudes within an $r=0\farcs2$ circular aperture.}
of ${\rm AB}=30$~mag in $RZYJH$, 29~mag in $F$, and 28~mag in $K$, which far 
exceed the ``deep/ultra-deep'' parts of the Core Community Surveys (CCS; see
Table~\ref{tab:depths}). 
To enable time-domain science, \acron\ will be split into three main epochs with a $\sim$1-yr cadence (the 3$^{\rm rd}$ epoch could be in Year 3 of mission operations, depending on scheduling), with each epoch further divided into three sub-epochs $\sim$10 days apart. \acron\ will enable unique studies of galaxies from $z\approx 14$ to $z< 2$, probe AGN to unprecedented low luminosities, and hunt for faint transients out to $z\gtrsim 5$.

\begin{table}[hbt!]
\setlength{\tabcolsep}{4pt}
\centering
\begin{tabular}{cc c c c c c c } \hline\hline
 & F062 ($R$) & F087 ($Z$) & F106 ($Y$) & F129 ($J$) & F158 ($H$) & F184 ($F$) & F213 ($K$) \\ \hline
 {\bf \acron} & 30.6 hrs & 53.5 hrs & 55.4 hrs & 65.4 hrs & 73.2 hrs & 36.5 hrs & 60.4 hrs\\
 5$\sigma$, $r=0.2$\arcsec & 30.0 mag & 30.0 mag& 30.0 mag & 30.0 mag & 30.0 mag& 29.0 mag & 28.0 mag \\
 \hline
 HLTDS stack & --- & 7.7 hrs & 5.9 hrs & 6.1 hrs & 8.4 hrs & 37.2 hrs & ---\\
 5$\sigma$, $r=0.2$\arcsec &  --- & 28.75 mag & 28.68 mag & 28.63 mag & 28.77 mag & 28.96 mag & --- \\
 \hline
 HLWAS D+UD & --- & 1.23 hrs & 3.68 hrs & 3.68 hrs & 3.68 hrs & 1.23 hrs & 1.23 hrs\\
 5$\sigma$, $r=0.2$\arcsec &  --- & 27.7 mag & 28.2 mag & 28.2 mag & 28.1 mag & 27.0 mag & 25.9 mag \\
 \hline
\end{tabular}
\label{tab:depths}
\caption{Comparison of integration times and depths in \acron\ and the two Community Core Surveys, the HLTDS 2-year stack and the HLWAS ``Deep'' (D) and ``Ultra-deep'' (UD). }
\end{table}

\section{Rationale}

    The iconic Hubble Ultra Deep Field (HUDF; \citealt{Beckwith2006}), 
subsequently observed with more programs and reprocessed as the Hubble eXtreme 
Deep Field (HXDF; \citealt{Oesch2010,Ellis2013,Illingworth2013,Teplitz2013}), 
invested $\sim$520 hours to reach ${\rm AB}\sim 29.7$$-$30.0~mag 
in most filters across $\sim$0.4$-$1.6\,$\mu$m with ACS and WFC3/IR over 
4.7~arcmin$^2$. The deepest JWST imaging has likewise invested $\sim$360 hours to reach 
${\rm AB} \sim 30.1$$-$30.9~mag over $\sim$9~arcmin$^2$ 
\citep{Eisenstein2025,Eisenstein2026}. 
But significantly extending these small areas at such depths is infeasible with 
these telescopes. With Roman's Wide Field Instrument (WFI), \acron\ will reach 
${\rm AB} \sim 30$~mag over $\sim$144$\times$ larger area than the full-depth 
ACS+WFC3/IR HXDF, revolutionizing studies of the deep universe.

$\bullet$ {\bf Why 7 Bands?}\,\,  The new science goals demand full Roman 
wavelength coverage with sufficient sampling for SED analysis (e.g., 
\citealt{Bagley2026}). In particular, $R$ (F062) is outside the range of JWST, while Euclid lacks $K$ (F213). Both are critical in making \acron\ unique among deep surveys. 

$\bullet$ {\bf Why Now?}\,\, Roman's first two cycles will be within Euclid's 
last three years and in the middle of its last $\sim$40 cadenced visits in the 
EUDF. Executing \acron\ early will ensure direct synergy with Euclid time-domain 
science. Furthermore, some of our discoveries will require prompt JWST follow-up at 
$>$2~$\mu$m; an early synergy with JWST will maximize the science returns from 
both observatories.

\begin{figure}
    \centering
    \includegraphics[width=\textwidth]{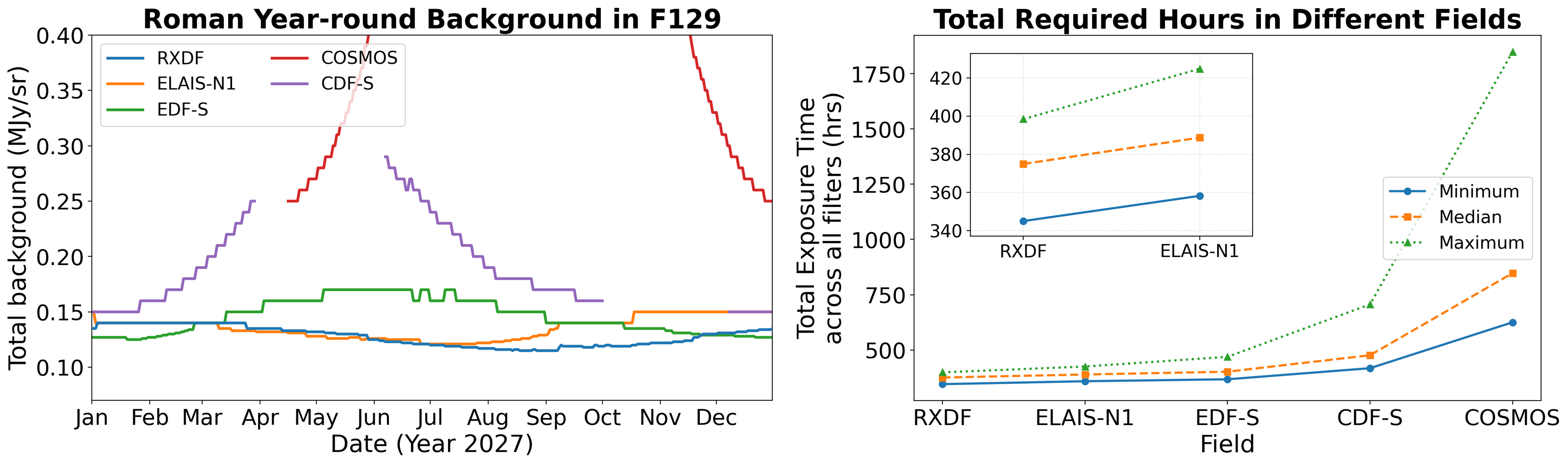}
    \caption{{(\bf left}) Total backgrounds in various extragalactic fields in their respective observing windows throughout a year, using F129 as example. 
({\bf right}) Total number of hours needed in these fields to reach our desired depths, which are affected by the changing zodiacal background in a year. For each field, the minimum (blue), maximum (green) and median (orange) number of hours are shown. The inset shows the comparison between \acron\ and ELAIS-N1.}
    \label{fig:bkexphrs}
\end{figure}

$\bullet$ {\bf Why choose this NEP field?}\,\, Besides the synergy with the EUDF 
and a myriad of deep data from radio to X-rays, our field has the lowest zodiacal 
background year-round compared to other fields and is thus the most suitable 
for creating an ultra-deep field with high efficiency (Figure~\ref{fig:bkexphrs}).
The NEP region is within the continuous viewing zone (CVZ) of Roman, which makes the 
\acron\ highly flexible in terms of scheduling.
Furthermore, the field is also within JWST's CVZ, and our high-value targets 
(especially transients) can be promptly observed by JWST at any time.
It already has JWST NIRCam imaging (F090W, F115W, F150W, F200W, F356W, F444W) 
covering $\sim$400~arcmin$^2$ to 5$\sigma$ limits of AB $=$ (26.9, 26.8, 26.8, 
28.0, 27.6, 27.8)~mag from the JWST NEXUS program. The NEP field has critical long-term 
strategic value for extragalactic science in the era of space survey facilities. 
Synergies with other facilities are further discussed below.

\begin{figure}
\centering
\includegraphics[width=\textwidth]{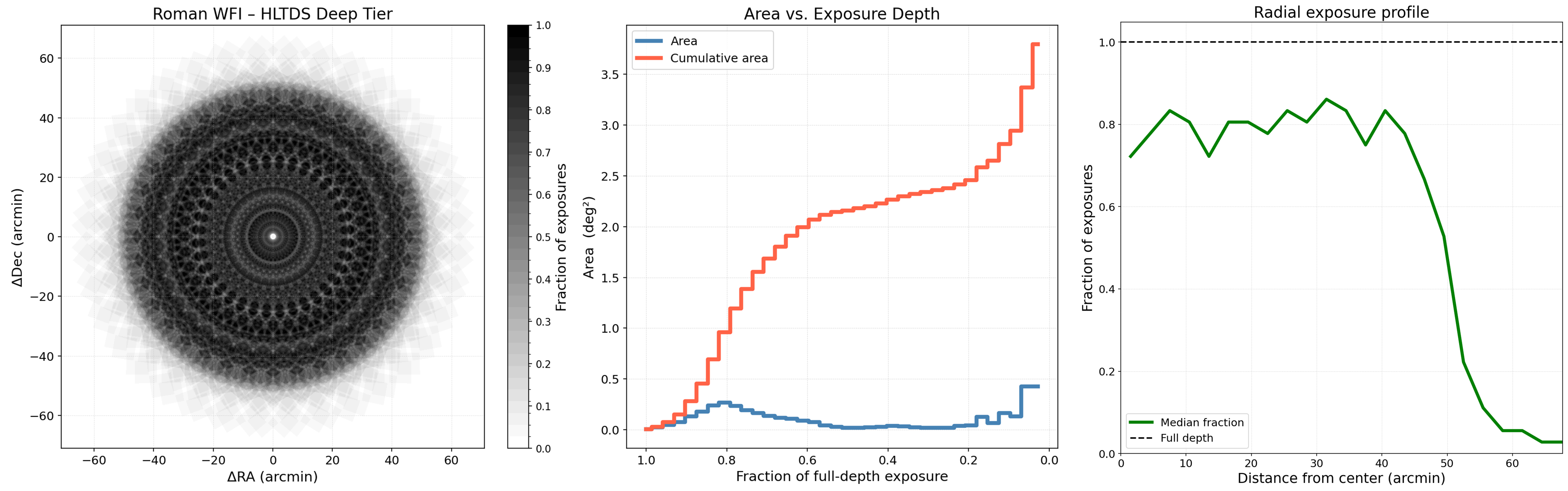}
\caption{
({\it left}) Exposure map of the 2-yr HLTDS stack, demonstrating its severe non-uniformity. The scale bar indicates the fractional integration with respect to the full-depth exposure.\,
({\it middle}) Differential (blue) and cumulative (red) areas as functions of fractional integration time.\,
({\it right}) Radial profile of the exposure map, calculated in annuli of 3~arcmin width. 
}
\label{fig:hltds}
\end{figure}

$\bullet$ {\bf Can it be done in a Roman CCS field to save time?}\,\,\, No. The 
most favorable alternative would be ELAIS-N1, which was chosen by the High Latitude
Time Domain Survey (HLTDS) as its northern field for its slightly lower Galactic 
reddening (E(B-V)$=$0.01~mag) than NEP, and this is important for the main science of 
the HLTDS on low-$z$ type Ia supernovae (SNe~Ia). However, ELAIS-N1 still has
significantly higher background for $\sim$5 months of the year, and avoiding this would put substantial
pressure on observatory scheduling. Moreover, even if we put the RXDF on top of the 
HLTDS 2-yr stack, it would not become a ``time-saver'' for two reasons. Firstly,
the HLTDS stack would only be equivalent to a small fraction of the time that \acron\ 
will need. Due to the short exposures used, the HLTDS is noticeably impacted by 
readout noise in most filters except in F184. The full-depth exposure times listed in 
Table~\ref{tab:depths} would be equivalent to 
($Z, Y, J, H, F$) = (5.4, 4.9, 5.2, 7.6, 33.9) hrs in our
adopted long-integration mode, and the possible time saving would not offset the 
cost due to its higher background. Secondly, more subtle but as important,
the HLTDS observing strategy will not fully cover the wide detector gaps and will
have severely non-uniform integrations across the stack (see 
Figure~\ref{fig:hltds}). Its full-depth area would amount to only 
21.1~arcmin$^2$ and would be scattered over the field. For a representative WFI pointing at 
its field center, the typical HLTDS coverage would only have $\sim$80\% of the 
full depth, which further negates any possible time saving.

\section{Observational Design}

\begin{figure}
    \centering
    \includegraphics[width=\textwidth]{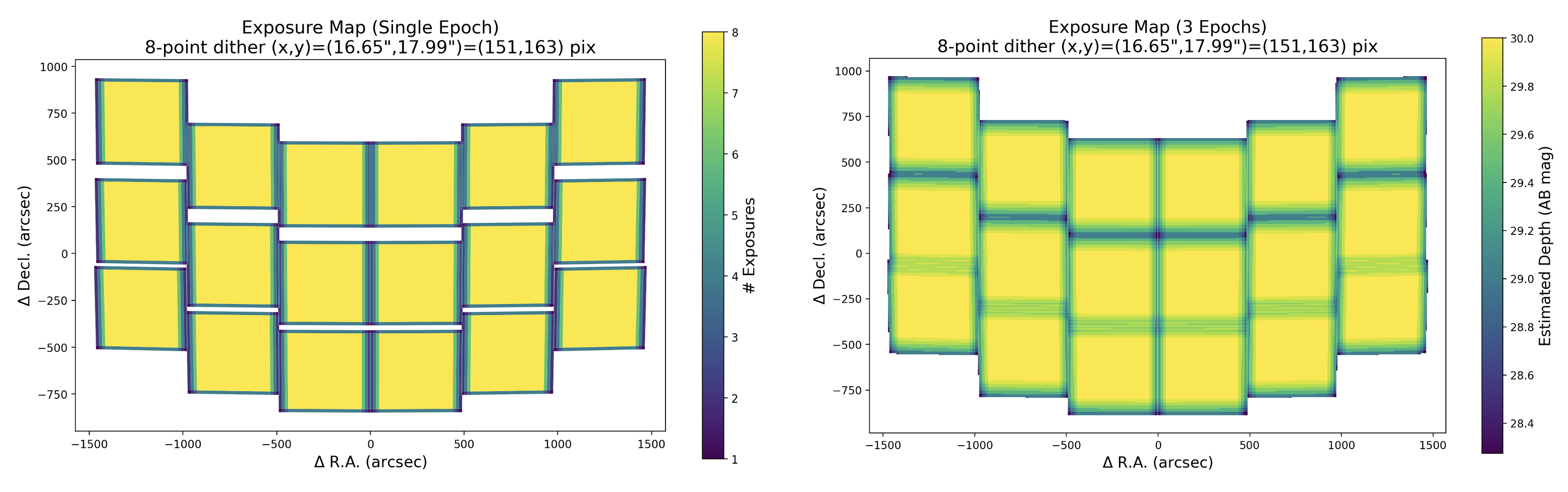}
    \caption{({\bf left}) \acron\ exposure map in a single sub-epoch (there being three 
    sub-epochs in each main epoch), where the gaps between detectors are 
    partially filled by our customized 8-point dithering pattern. The color bar 
    scale indicates the number of overlapped exposures. ({\bf right}) Exposure 
    map of the 3-yr stack made out of three main epochs (nine sub-epochs). The 
    gaps are filled by moving the field center slightly in between the 
    sub-epochs. The color bar scale indicates the 5$\sigma$ depth in AB mag. The 
    full-depth area amounts to 678.75~arcmin$^2$. The total, contiguous coverage 
    is 1,243~arcmin$^2$.    
    }
    \label{fig:expmap}
\end{figure}

    We aim to maximize depth over the full WFI footprint and will split the observations into 3 main
epochs separated by $\sim$1 yr. As the telescope's orientation rotates on a 
yearly basis, such a yearly cadence will ensure the same WFI FoV being repeated 3 
times to maximize the overlapping (full-depth) area. Each main epoch is further 
divided into 3 sub-epochs separated by $\sim$10$-$12 days, with position angle 
adjustments (allowed within $\pm 15^{\rm o}$ at Roman) to maintain the same
footprint. As the exposure maps in Figure~\ref{fig:expmap} show, we will use a 
customized dithering pattern to cover the gaps between detectors so that they are 
all filled in the 3-yr stack while maximizing the full-depth area.

\begin{figure}
    \centering
    \includegraphics[width=\textwidth]{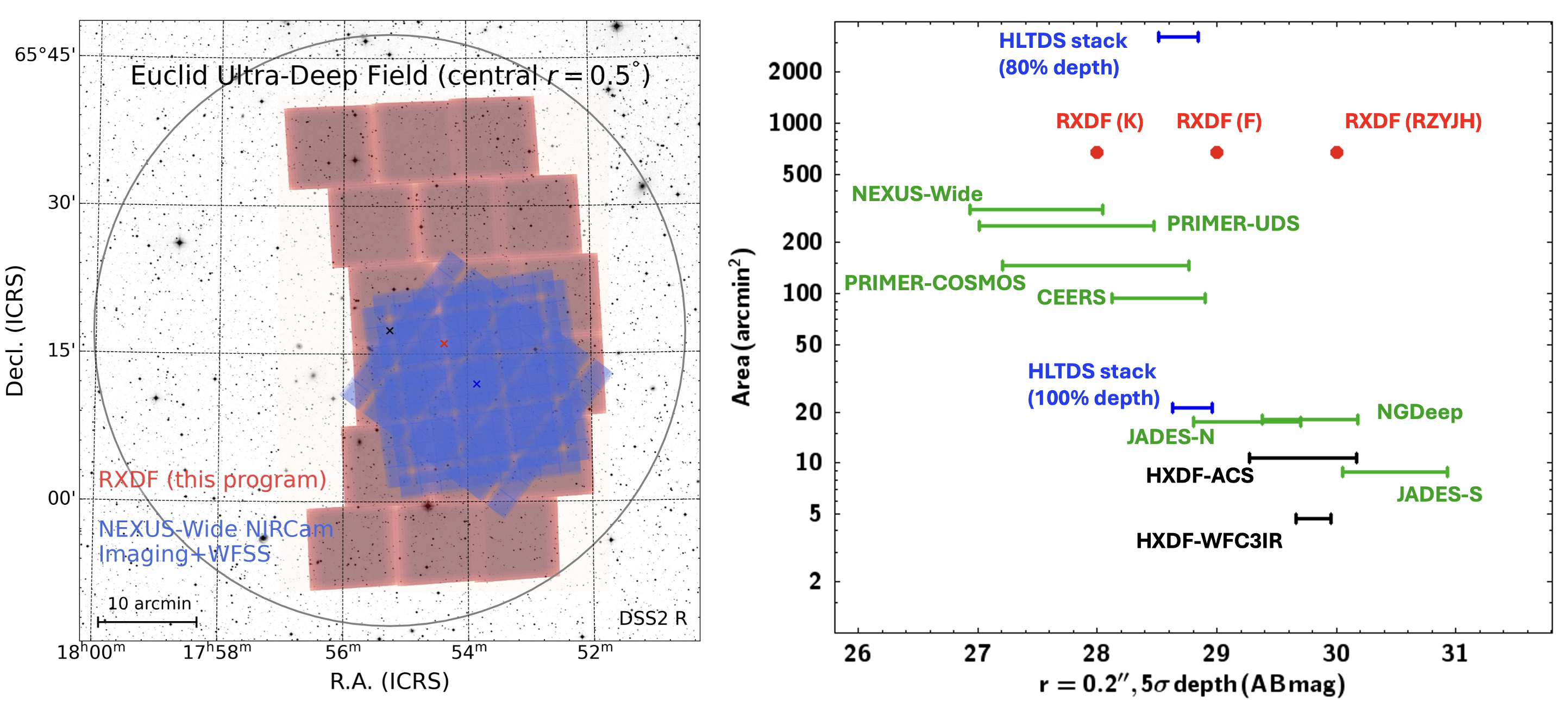}
    \caption{({\bf Left}) Overlay of \acron\ on the center of EUDF (black circle) 
    and its overlap with the JWST-NEXUS field (blue), at a representative orient. Regardless of the exact execution date, the \acron\ 
    footprint placement can always be chosen to be within the EUDF and to 
    cover the JWST-NEXUS field.
    \acron\ will reach $\sim$2.5~mag deeper than JWST-NEXUS 
    over 0.8--1.7~$\mu$m and extend bluer to 0.5~$\mu$m.
    ({\bf Right}) Depth and coverage of \acron\ (red circles) as compared to
    HXDF (black), various JWST surveys to date (green) and Roman's HLTDS (blue).
    The depths of these other surveys are indicated by horizontal bars to reflect
    the different depths achieved in different bands involved.
    }
    \label{fig:fov_depthcomp}
\end{figure}

   As demonstrated by numerous studies, proper SED analysis relies on a sufficient 
number of passbands covering the full range from optical to near-IR wavelengths (e.g., 
\citealt{Illingworth2013, Bagley2026}). We stress that neither $R$ or $K$ can be 
omitted: $R$ will be critical for the studies not only at mid-$z$ but also at 
high-$z$ (\citealt{Bagley2026}), while $K$, even by just providing upper limits, 
will be crucial in distinguishing SEDs of AGNs and normal galaxies. 

   To minimize the impact of read-out noise, we will use $>$600~sec integration 
time. We adopt the 5-$\sigma$ depth measured within an $r=0\farcs2$ aperture for 
point sources, which is the Roman ETC's default for depth calculation. For each 
day within a cycle, we calculate the exposure times needed to reach the desired 
depths, and the median values are listed in  Table~\ref{tab:depths}. The total
on-source time is 375 hrs; after taking the overhead into account (as calculated by
the APT), the total charged time is 386.41 hrs, which is what our program has been
approved for.

   The \acron\ footprint within the EUDF is demonstrated in the left panel of 
Figure~\ref{fig:fov_depthcomp}, at a representative orientation. The right panel of
this figure compares the \acron\ depth and coverage to those of the HXDF, 
various JWST surveys to date and Roman's HLTDS.

\begin{figure}
\centering
\includegraphics[width=\textwidth]{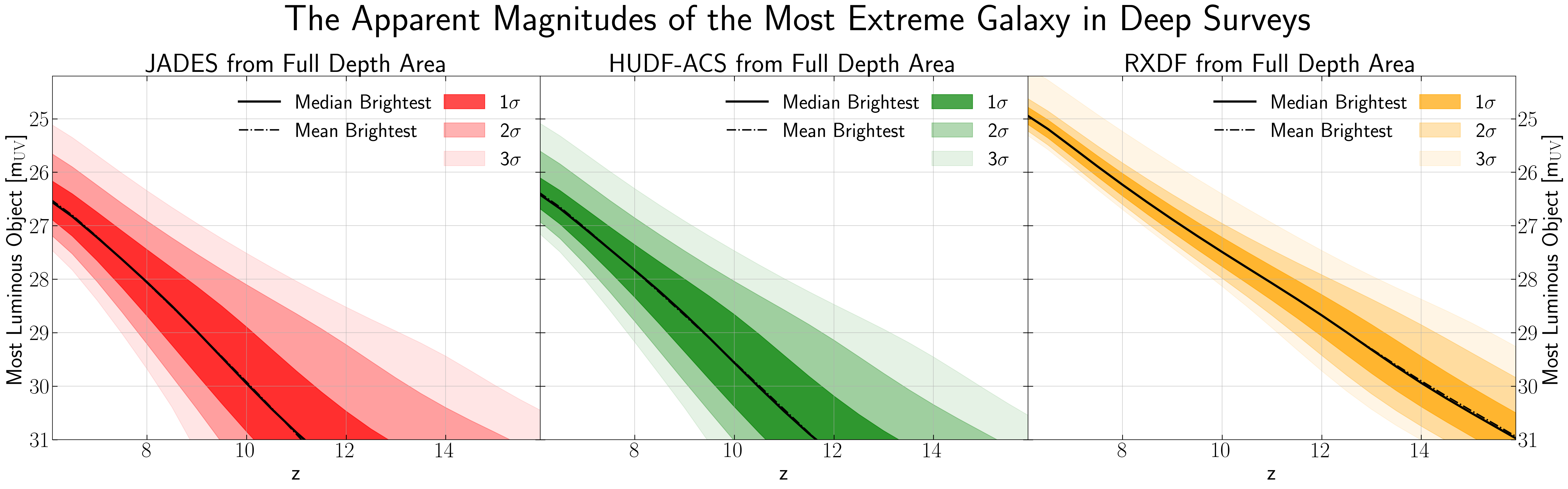}
\caption{Impact of cosmic variance on characterizing rare objects due to the size of survey area, using the brightest galaxies at high-$z$ as example. The probability contours of the brightest galaxy that could be found in areas of JADES, HUDF and \acron\ (calculated following \citealt{Jespersen2025}) are shown from left to right. \acron's large area increases the chance of finding such objects and greatly reduces the impact of cosmic variance when calculating their true number density. 
}
\label{fig:cvdemo}
\end{figure}

\section{Key Science Goals}

  \acron\ will create a full-depth area of 678.75~arcmin$^2$, which is embedded 
in a contiguous coverage of 1,243~arcmin$^2$. The unprecedented combination of 
depth and area will minimize the impact of cosmic variance for many studies.
Figure~\ref{fig:cvdemo} demonstrates this using the brightest galaxies
at high-$z$ as examples. We lay out below the unique science questions that only 
\acron\ can properly address. 

\subsection{Galaxies}

   JWST has pushed the redshift frontier of galaxies to $z=14.44$ 
\citep{Naidu2026} and is likely to soon break the record again. These 
discoveries have raised numerous crucial questions at $z\lesssim 14$, and we 
highlight below a few key topics that can only be properly addressed by \acron. 

\subsubsection{Cosmic dawn}

Candidate galaxies can be selected as dropouts in successive bands from $Z$ to $F$, reaching 30, 29 and 28 mag for those at $z\lesssim 10$, $z\approx 12$ and $z\approx 14$, respectively, which correspond to $M_{\rm UV}= -16.8$, $-17.1$, $-17.5$, $-18.8$, and $-20.0$~mag at $z\approx 6.2$, 7.7, 9.6, 12.0, and 14.1, respectively (Figure 2). By conservative estimates and after considering incompleteness, our sample will contain $\sim$$10^4$, $\sim$$5\times 10^3$, $\sim$$10^3$, $\sim$100 and $\sim$20 galaxies in these redshift ranges.

   (1) {\it Galaxy luminosity function (LF):}\,\, Accurate measurements of galaxy LFs at $z\gtrsim 6$ have profound impact on our knowledge of early galaxy formation and reionization. \acron's results will be decisive at both the bright and faint ends for the $z\lesssim 11$ LFs. At $z\gtrsim 12$, current LFs are highly uncertain due to small number statistics, but \acron\ will impose the strongest constraints.

   (2) {\it Constraining modes of star formation and host halo masses with clustering:}\,\, One of JWST's biggest surprises was the discovery of many more bright galaxies at $z>10$ than expected. The two leading explanations are a higher overall star formation efficiency or more bursty star formation at earlier times \citep{Somerville2025}. These mechanisms have degenerate effects on the UVLF but make very different predictions for the host halo masses, $M_h$, that galaxies of a given luminosity occupy. 
   Because halo clustering is a strong function of mass, the simultaneous measurements of LF and galaxy clustering can not only constrain the $M_h$-$L_{\rm UV}$ relationship but also the timescale of SF burstiness \citep{lee09,Endsley2020,Munoz2023,Sun2025}.

   As \acron\ spans a large and contiguous area with largely uniform depth, we can measure the angular correlation function (ACF; e.g., \citealt{LandySzalay1993}) 
   or counts-in-cells (field to field variance) 
   across a broad range of luminosities, from bright objects inaccessible to pencil-beam surveys like JADES to faint objects missed by shallow wide-field surveys like COSMOS-Web \citep[as discussed for example by ][]{Paquereau2025, Dalmasso2026}. These measurements will yield accurate constraints on the relationship between UV luminosity and host halo mass, 
   discriminate between different theoretical scenarios to explain the ``excess'' bright galaxies.

\begin{figure}
\centering
\includegraphics[width=\textwidth]{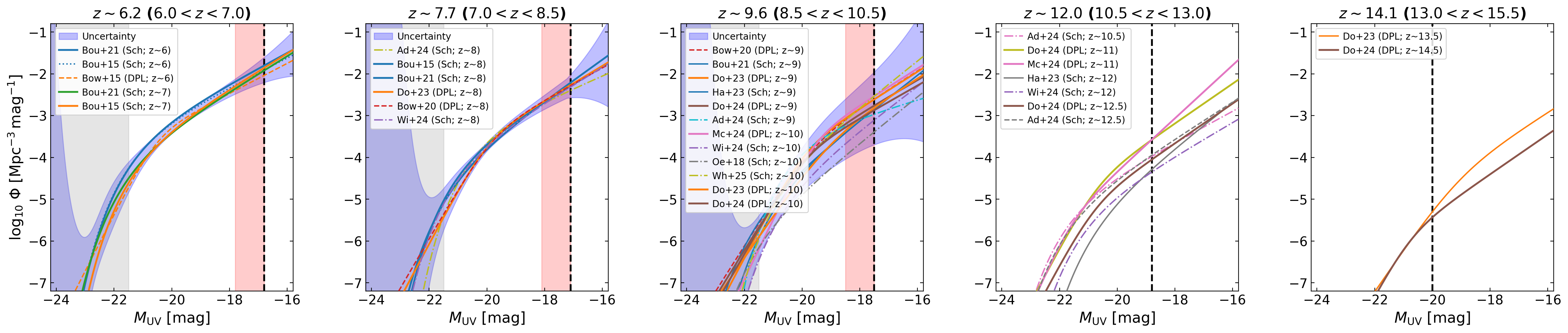}
\caption{Current estimates of UVLF at $6\lesssim z\lesssim 14$, taken from 
\cite{Adams2024, Bouwens2015, Bouwens2021, Bowler2015, Bowler2020, Donnan2023, Donnan2024, Harikane2023a, McLeod2024, Oesch2018, Whitler2025, Willott2024}. 
The \acron\ limits are shown by the vertical dashed lines. The blue area in the first three panels indicates the 1$\sigma$ uncertainty, while the grey and red shaded regions indicate the regimes where \acron\ will provide decisive results. The LFs at $z\gtrsim 12$ are highly uncertain, and our results will impose the strongest constraints. All this is owing to the minimal cosmic variance in \acron\ as compared to any other (including JWST) surveys.
}
\end{figure}

\begin{figure}
\centering
\includegraphics[width=\textwidth]{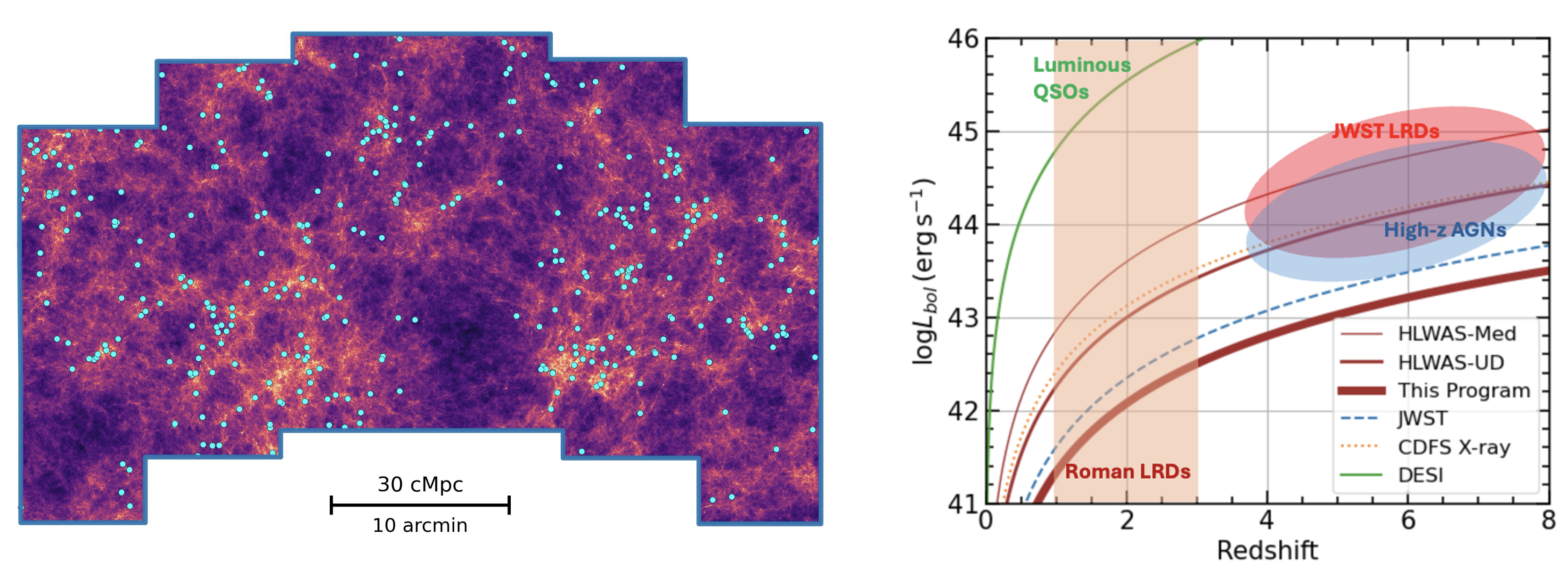}
\caption{({\bf left}) Demonstration of early LSS from the TNG300 simulation following \cite{WuZihao2026}. A slice at $z=10$ with $\Delta z = 0.2$ (40 cMpc) is shown in the \acron\ footprint. Galaxies of rest-frame $m_{\rm UV}<30$ mag are indicated in cyan and show obvious LSS. Only \acron\ can beat down the cosmic variance and properly interpret such early structures. 
({\bf right}) Bolometric luminosity and redshift space for AGNs and LRDs. \acron\ will deliver the deepest AGN sample at $z\gtrsim 6$, on par with the deepest JWST AGN samples but with $\gtrsim$30--50$\times$ larger size. \acron\ also probes cosmic-noon LRDs to $\sim$10$\times$ lower luminosity than high-$z$ JWST LRDs. 
}
\label{fig:lss_agn}
\end{figure}

   (3) {\it Early large scale structures (LSS):}\,\,  
   The emergence of LSS at early epochs provides a critical constraint on the hierarchical formation theory. While JWST has revealed a few structures detected as galaxy overdensities as early as 
   $z\approx 10.5$ \citep{Tacchella2023b, Scholtz2024, WuZihao2026},  
   the abundance of such massive structures and how they develop over cosmic time remain unconstrained.
   Properly addressing this question requires a deep survey limit sufficient to detect low-luminosity members over a wide, contiguous area minimizing cosmic variance (Figure~\ref{fig:lss_agn}, left). \acron\ will, for the first time, offer a unique opportunity to conduct a systematic LSS study over $6\lesssim z\lesssim 14$. 

\subsubsection{Cosmic noon to afternoon}

The epoch of $z\approx 2$--3 (cosmic noon) marks the peak of cosmic SFR density. 
After this peak, prevalent quenching leads to a rapid buildup of quiescent populations, especially in the low-mass regime \citep{muzzin2013}. \acron\ will detect dwarf galaxies down to $M_*\sim 10^{7.0}, 10^{7.5}, 10^{8.0} M_\odot$ at $z\sim 1, 2, 3$ while being complete for the oldest passive galaxies to $M_*\sim 10^{8.0}, 10^{8.5}, 10^{9.0} M_\odot$ at these redshifts (estimated following \citealt{Pozzetti2010,Chartab2020}). Our depth and area will enable an unprecedented study of dwarf galaxies and their environments.

Environmental effects are believed to be the primary reason of quenching in low-mass galaxies at $z<1$ \citep[e.g.,][]{geha12,wetzel2013,balogh16,fossati17,ycguo17}. However, the role of environmental quenching at higher redshift remains unclear due to scarce observations. Based on the predicted Roman $z_{\rm phot}$ accuracy \citep{romanphotoz}, our test shows that the $7^{th}$-neighbor overdensity measured by \acron\ at $1<z<3$ will recover local densities with a scatter of $0.25$ dex. This is significantly smaller than the full dynamic range of environments \citep{poggianti2010}, allowing us to statistically distinguish between voids, filaments, and clusters. At $1 < z < 3$, \acron\ will robustly: (1) detect low-mass quiescent galaxies, (2) quantify quenching fractions across mass and environment, and (3) correlate galaxy morphology with local density. \acron\ will provide the first comprehensive view of the multi-dimensional interplay between stellar mass, star formation, morphology, and environments in the dwarf regime at cosmic noon and afternoon and hence offer the most stringent constraints to date on early galaxy quenching.

\subsection{Accreting supermassive black holes (SMBHs)}

Identifying low-luminosity AGNs and measuring their physical properties towards cosmic dawn provide crucial insights on early SMBH growth and their coevolution with galaxies. AGN selection at the extreme faint-luminosity is challenging: contamination from host galaxy and photometric errors prevent robust AGN identification using traditional SED-based approaches \citep{Lyu2024}. \acron\ will efficiently identify unobscured AGNs up to $z\gtrsim 6$ via sensitive variability detection over two years. 
AGNs vary more strongly in rest-frame UV than in longer wavelengths; our bands sample rest-frame UV at $z\gtrsim 6$ and therefore will be more sensitive than JWST in identifying early AGNs \citep{Stone2025,Scholtz2025}.
The variability selection can also identify extremely faint AGNs at all redshifts, including dwarf AGNs ($\lesssim 10^{5-6}\,M_\odot$) at cosmic noon far beyond the reach of other programs. Our sample will probe $\sim$10$\times$ fainter than those from the HLWAS Ultra-Deep, reaching similar depths but considerably larger sample sizes compared with the deepest X-ray and JWST programs \citep[][]{Scholtz2025}. This is illustrated in the right panel of Figure~\ref{fig:lss_agn}. Such a faint AGN sample would lead to a leap-forward in our understanding of SMBH formation, e.g., $L_{\rm bol}\approx 10^{42}\,{\rm erg\,s^{-1}}$ at $z\sim 2$ corresponds to a $10^4\,M_\odot$ (seed) black hole accreting at the Eddington limit.   

\acron\ will also enable crucial advances in studying Little Red Dots (LRDs), an enigmatic population discovered by JWST \citep{Labbe2023, Kocevski2023, Harikane2023b, Matthee2024, Greene2024}, with compact morphology, V-shaped SEDs, and abundant at high-$z$. Their nature is debated, including dusty star formation, extreme accretion, and/or gas enshrouded accreting SMBHs \citep{Akins2025,Inayoshi2025,lambrides25,Naidu2025,LiuH2025} that may hold the key to SMBH seed formation. Addressing this requires vastly expanding the samples at the faint end, uniquely achieved with \acron\ pushing to luminosities $\sim$10$\times$ below the current JWST samples (with Roman LRDs at $z$$\sim$1$-$3) and $>$3 mag below the least luminous LRDs in the HLWAS Ultra-Deep. Our large area is necessary to yield the first statistically large samples at these uncharted low luminosities given their current estimated (albeit highly uncertain) space densities at Cosmic Noon \cite[c.f.][]{Ma2025}. Moreover, their variability \citep{Stone2025, Burke2025} will be probed to unparalleled depths. Our data will uniquely enable vast improvements in constraining and testing physical models of LRDs \citep{Inayoshi2025,LiuH2025,Naidu2025,Secunda2026}.

\subsection{High-$z$ supernovae}

In our cadenced observations, each sub-epoch will reach ${\rm AB} = 28.8$~mag in $RZYJH$, 27.8~mag in $F$ and 26.8~mag in $K$, and each main epoch (stacks of 3 sub-epochs) will reach  $RZYJH$ = 29.4, $F$ = 28.4 and $K$ = 27.4~mag.
These unprecedented sub-/single-epoch depths will probe uncharted parameter space for transient discovery, especially for high-$\mathbf{z}$ SNe. We will easily detect SNe at $z\geq 2.5$, complementing the HLTDS whose focus is SNe at $z<2.5$. Our layered time cadences and multi-band light curves, aided by host galaxies' superb $z_{\rm phot}$, will allow  SN classification with high confidence. 

   (1) {\it Thermonuclear SNe, or SNe~Ia:}\,\, SNe~Ia are currently the only standard candles extending the Hubble diagram to high-$z$ and therefore are critical to precision cosmology. SNe~Ia need white dwarf progenitors and cannot occur too early, however,
   it is believed that SNe~Ia could begin at $z\sim 5$ \citep[e.g.,][]{RiessLivio2006, Graur2011, Wiseman2021}. With peak of $M\approx -19.3$ ($m=27.1$ at $z=5$), they can be easily detected \emph{in a sub-epoch} (but missed by the HLTDS). We predict $\sim$200 SNe~Ia \emph{in each epoch}, of which $\sim$40 are at $z>2.5$. The sub-epochs will well sample the light curves in the first 30 days of discovery to robustly select those near peak (expecting 6--22 of them) to enable cosmological applications.
   
   (2) {\it Core-collapse SNe (CCSNe):}\,\, As the progenitors of CCSNe are high mass ($\gtrsim 8M_\odot$) stars, the rate of CCSNe constrains the high-mass end of their host galaxies' IMF, which is fundamental to understanding galaxy evolution. CCSNe can be detected out to $z\approx 8$ as transient Y-dropouts in our single-epoch images when near their peak ($M\approx -18.0$~mag, or $m\approx 29.2$~mag). We expect $\sim$400 CCSNe at $z>2.5$ and $\sim$10 at $z>6$.

   (3) {\it Superluminous SNe (SLSNe):}\,\, With peak luminosity $M$$<$$-21$~mag, SLSNe can be detected out to $z\approx 14$ as transient $F$-dropouts ($K$=27 for $M$=$-21$ mag) in our single-epoch images. The 1-yr cadence corresponds to rest-frame $\sim$24 days at $z\approx 14$, which will provide sufficient time sampling. Based on the $z\sim 2$--5 observations to date \citep{Cooke2012, Prajs2017}, \acron\ will detect $\sim$80 SLSNe at $z>2.5$. The progenitors of SLSNe must be very high-mass stars, and it is believed that pair instability could be a viable mechanism for SLSNe. Pop-III (i.e., metal-free) stars could end their lives as supernovae through pair instability, and therefore it is likely that the SLSNe found at the highest redshift from \acron\ could contain Pop-III SNe.

\subsection{Broader science}

   \acron\ will enable a plethora of  studies beyond the above, even extending to the studies of our Galaxy and the Solar System. Two examples are outlined below.

   --- Galactic cool dwarfs:\,\, Using colors, PSF shapes and proper motions, \acron\ will identify and characterize the lowest-mass stars and brown dwarfs (BDs; $<$0.1~$M_\odot$) through the thin and thick disks and deep into the halo. We will (a) measure thin disk scale height vs. temperature across the BD regime to examine evolutionary 
   ages and mass-dependent scattering in the Galactic potential \citep{Ryan2017, Aganze2022}; (b) measure the LF and MF of the thick disk and halo across the star/BD limit for the first time; (c) build a large sample of halo BDs ($>$100)
   to explore abundance effects on low-temperature atmospheres \citep{Burgasser2025} and thermal evolution across the star/BD gap; and 
   (d) identify the lowest-mass stars of the outer halo ($>$100 kpc) 
   and determine their kinematic alignment with stream populations.

   --- Moving objects in our Solar system:\,\, The Oort cloud comets (OCCs) have a significant chance of being near the ecliptic poles \citep{KaibQuinn2009}, and each of our sub-epochs is ideal for their detection by shift-and-stack \citep[e.g.,][]{Bernstein2004, Smotherman2021}. At $R = 28.8$~mag for each sub-epoch, typical OCC objects ($\sim$200 meters) are reachable at 4~AU. Towards larger distances, it has been posited that there should be substantial numbers of dynamically old comets  beyond the orbit of Saturn \citep{Kaib2022}, but these are too faint to be detected by any current surveys. \acron's unprecedented area and depth will allow us to quantitatively test this hypothesis, placing stringent limits even in case of non-detection. 

\section{Tentative Data Release Plan}

   As per the general Roman data access policy, all the raw and Roman pipeline-processed data from 
our program will be immediately public. However, we anticipate carrying out some additional calibrations beyond the default pipeline, specific to the needs of our RXDF program, in order to optimize the data products 
for achieving the aforementioned science goals. We plan to release the team-vetted 
data products on an annual basis. To facilitate the science production by the
community at large, we will also make best-effort, ``quick releases'' of our
products described as follows.

    $\circ$\,\,  We plan to release the image products of each major epoch as soon as the validation is finished, which we expect will be one month after the completion of the observations for that major epoch. These include the images of each sub-epoch as well as the stacks of the three sub-epochs.

    $\circ$\,\,  The source catalogs will be released one month after the per-epoch image release. 

    $\circ$\,\,  The search for transients will be carried out as soon as each sub-epoch is finished, and the results will be immediately announced through the TNS to facilitate possible follow-ups.

    $\circ$\,\, The full-stacks of all data will be expected to be released one month after the last observation, and the accompanying source catalogs will also be released one month after the image release.



\newpage

\begin{thebibliography}{}
\expandafter\ifx\csname natexlab\endcsname\relax\def\natexlab#1{#1}\fi
\providecommand{\url}[1]{\href{#1}{#1}}
\providecommand{\dodoi}[1]{doi:~\href{http://doi.org/#1}{\nolinkurl{#1}}}
\providecommand{\doeprint}[1]{\href{http://ascl.net/#1}{\nolinkurl{http://ascl.net/#1}}}
\providecommand{\doarXiv}[1]{\href{https://arxiv.org/abs/#1}{\nolinkurl{https://arxiv.org/abs/#1}}}

\bibitem[{{Adams} {et~al.}(2024){Adams}, {Conselice}, {Austin}, {Harvey},
  {Ferreira}, {Trussler}, {Juod{\v{z}}balis}, {Li}, {Windhorst}, {Cohen},
  {Jansen}, {Summers}, {Tompkins}, {Driver}, {Robotham}, {D'Silva}, {Yan},
  {Coe}, {Frye}, {Grogin}, {Koekemoer}, {Marshall}, {Pirzkal}, {Ryan},
  {Maksym}, {Rutkowski}, {Willmer}, {Hammel}, {Nonino}, {Bhatawdekar},
  {Wilkins}, {Bradley}, {Broadhurst}, {Cheng}, {Dole}, {Hathi}, \&
  {Zitrin}}]{Adams2024}
{Adams}, N.~J., {Conselice}, C.~J., {Austin}, D., {et~al.} 2024, \apj, 965,
  169, \dodoi{10.3847/1538-4357/ad2a7b}

\bibitem[{{Aganze} {et~al.}(2022){Aganze}, {Burgasser}, {Malkan}, {Theissen},
  {Tejada Arevalo}, {Hsu}, {Bardalez Gagliuffi}, {Ryan}, \&
  {Holwerda}}]{Aganze2022}
{Aganze}, C., {Burgasser}, A.~J., {Malkan}, M., {et~al.} 2022, \apj, 934, 73,
  \dodoi{10.3847/1538-4357/ac7053}

\bibitem[{{Akins} {et~al.}(2025){Akins}, {Casey}, {Lambrides}, {Allen},
  {Andika}, {Brinch}, {Champagne}, {Cooper}, {Ding}, {Drakos}, {Faisst},
  {Finkelstein}, {Franco}, {Fujimoto}, {Gentile}, {Gillman}, {Gozaliasl},
  {Harish}, {Hayward}, {Hirschmann}, {Ilbert}, {Kartaltepe}, {Kocevski},
  {Koekemoer}, {Kokorev}, {Liu}, {Long}, {McCracken}, {McKinney}, {Onoue},
  {Paquereau}, {Renzini}, {Rhodes}, {Robertson}, {Shuntov}, {Silverman},
  {Tanaka}, {Toft}, {Trakhtenbrot}, {Valentino}, \& {Zavala}}]{Akins2025}
{Akins}, H.~B., {Casey}, C.~M., {Lambrides}, E., {et~al.} 2025, \apj, 991, 37,
  \dodoi{10.3847/1538-4357/ade984}

\bibitem[{{Bagley} {et~al.}(2025){Bagley}, {Finkelstein}, {Rhoads}, {Malhotra},
  {Yung}, {Somerville}, \& {Papovich}}]{Bagley2026}
{Bagley}, M., {Finkelstein}, S., {Rhoads}, J., {et~al.} 2025, arXiv e-prints,
  arXiv:2603.09828, \dodoi{10.48550/arXiv.2603.09828}

\bibitem[{{Balogh} {et~al.}(2016){Balogh}, {McGee}, {Mok}, {Muzzin}, {van der
  Burg}, {Bower}, {Finoguenov}, {Hoekstra}, {Lidman}, {Mulchaey}, {Noble},
  {Parker}, {Tanaka}, {Wilman}, {Webb}, {Wilson}, \& {Yee}}]{balogh16}
{Balogh}, M.~L., {McGee}, S.~L., {Mok}, A., {et~al.} 2016, \mnras, 456, 4364,
  \dodoi{10.1093/mnras/stv2949}

\bibitem[{{Beckwith} {et~al.}(2006){Beckwith}, {Stiavelli}, {Koekemoer},
  {Caldwell}, {Ferguson}, {Hook}, {Lucas}, {Bergeron}, {Corbin}, {Jogee},
  {Panagia}, {Robberto}, {Royle}, {Somerville}, \& {Sosey}}]{Beckwith2006}
{Beckwith}, S. V.~W., {Stiavelli}, M., {Koekemoer}, A.~M., {et~al.} 2006, \aj,
  132, 1729, \dodoi{10.1086/507302}

\bibitem[{{Bernstein} {et~al.}(2004){Bernstein}, {Trilling}, {Allen}, {Brown},
  {Holman}, \& {Malhotra}}]{Bernstein2004}
{Bernstein}, G.~M., {Trilling}, D.~E., {Allen}, R.~L., {et~al.} 2004, \aj, 128,
  1364, \dodoi{10.1086/422919}

\bibitem[{{Bouwens} {et~al.}(2015){Bouwens}, {Illingworth}, {Oesch}, {Trenti},
  {Labb{\'e}}, {Bradley}, {Carollo}, {van Dokkum}, {Gonzalez}, {Holwerda},
  {Franx}, {Spitler}, {Smit}, \& {Magee}}]{Bouwens2015}
{Bouwens}, R.~J., {Illingworth}, G.~D., {Oesch}, P.~A., {et~al.} 2015, \apj,
  803, 34, \dodoi{10.1088/0004-637X/803/1/34}

\bibitem[{{Bouwens} {et~al.}(2021){Bouwens}, {Oesch}, {Stefanon}, {Illing
  worth}, {Labb{\'e}}, {Reddy}, {Atek}, {Montes}, {Naidu}, {Nanayakkara},
  {Nelson}, \& {Wilkins}}]{Bouwens2021}
{Bouwens}, R.~J., {Oesch}, P.~A., {Stefanon}, M., {et~al.} 2021, \aj, 162, 47,
  \dodoi{10.3847/1538-3881/abf83e}

\bibitem[{{Bowler} {et~al.}(2020){Bowler}, {Jarvis}, {Dunlop}, {Mc Lure},
  {McLeod}, {Adams}, {Milvang-Jensen}, \& {M cCracken}}]{Bowler2020}
{Bowler}, R.~A.~A., {Jarvis}, M.~J., {Dunlop}, J.~S., {et~al.} 2020, \mnras,
  493, 2059, \dodoi{10.1093/mnras/staa313}

\bibitem[{{Bowler} {et~al.}(2015){Bowler}, {Dunlop}, {McLure}, {McCracken},
  {Milvang-Jensen}, {Furusawa}, {Taniguchi}, {Le F{\`e}vre}, {Fynbo}, {Jarvis},
  \& {H{\"a}u{\ss}ler}}]{Bowler2015}
{Bowler}, R.~A.~A., {Dunlop}, J.~S., {McLure}, R.~J., {et~al.} 2015, \mnras,
  452, 1817, \dodoi{10.1093/mnras/stv1403}

\bibitem[{{Burgasser} {et~al.}(2025){Burgasser}, {Schneider}, {Meisner},
  {Caselden}, {Hsu}, {Gerasimov}, {Aganze}, {Softich}, {Karpoor}, {Theissen},
  {Brooks}, {Bickle}, {Gagn{\'e}}, {Artigau}, {Marsset}, {Rothermich},
  {Faherty}, {Kirkpatrick}, {Kuchner}, {Andersen}, {Beaulieu}, {Colin},
  {Gantier}, {Gramaize}, {Hamlet}, {Hinckley}, {Kabatnik}, {Kiwy}, {Martin},
  {Massat}, {Pendrill}, {Sainio}, {Sch{\"u}mann}, {Th{\'e}venot}, {Walla},
  {W{\k{e}}dracki}, \& {Backyard Worlds: Planet 9
  Collaboration}}]{Burgasser2025}
{Burgasser}, A.~J., {Schneider}, A.~C., {Meisner}, A.~M., {et~al.} 2025, \apj,
  982, 79, \dodoi{10.3847/1538-4357/adb39f}

\bibitem[{{Burke} {et~al.}(2025){Burke}, {Stone}, {Shen}, \&
  {Jiang}}]{Burke2025}
{Burke}, C.~J., {Stone}, Z., {Shen}, Y., \& {Jiang}, Y.-F. 2025, arXiv
  e-prints, arXiv:2511.16082, \dodoi{10.48550/arXiv.2511.16082}

\bibitem[{{Chartab} {et~al.}(2020){Chartab}, {Mobasher}, {Darvish},
  {Finkelstein}, {Guo}, {Kodra}, {Lee}, {Newman}, {Pacifici}, {Papovich},
  {Sattari}, {Shahidi}, {Dickinson}, {Faber}, {Ferguson}, {Giavalisco}, \&
  {Jafariyazani}}]{Chartab2020}
{Chartab}, N., {Mobasher}, B., {Darvish}, B., {et~al.} 2020, \apj, 890, 7,
  \dodoi{10.3847/1538-4357/ab61fd}

\bibitem[{{Cooke} {et~al.}(2012){Cooke}, {Sullivan}, {Gal-Yam}, {Barton},
  {Carlberg}, {Ryan-Weber}, {Horst}, {Omori}, \& {D{\'\i}az}}]{Cooke2012}
{Cooke}, J., {Sullivan}, M., {Gal-Yam}, A., {et~al.} 2012, \nat, 491, 228,
  \dodoi{10.1038/nature11521}

\bibitem[{{Dalmasso} {et~al.}(2026){Dalmasso}, {Ferrami}, {Leethochawalit},
  {Ventura}, \& {Trenti}}]{Dalmasso2026}
{Dalmasso}, N., {Ferrami}, G., {Leethochawalit}, N., {Ventura}, E.~M., \&
  {Trenti}, M. 2026, \mnras, 546, stag001, \dodoi{10.1093/mnras/stag001}

\bibitem[{{Donnan} {et~al.}(2023){Donnan}, {McLeod}, {Dunlop}, {McLure},
  {Carnall}, {Begley}, {Cullen}, {Hamadouche}, {Bowler}, {Magee}, {McCracken},
  {Milvang-Jensen}, {Moneti}, \& {Targett}}]{Donnan2023}
{Donnan}, C.~T., {McLeod}, D.~J., {Dunlop}, J.~S., {et~al.} 2023, \mnras, 518,
  6011, \dodoi{10.1093/mnras/stac3472}

\bibitem[{{Donnan} {et~al.}(2024){Donnan}, {McLure}, {Dunlop}, {McLeod},
  {Magee}, {Arellano-C{\'o}rdova}, {Barrufet}, {Begley}, {Bowler}, {Carnall},
  {Cullen}, {Ellis}, {Fontana}, {Illingworth}, {Grogin}, {Hamadouche},
  {Koekemoer}, {Liu}, {Mason}, {Santini}, \& {Stanton}}]{Donnan2024}
{Donnan}, C.~T., {McLure}, R.~J., {Dunlop}, J.~S., {et~al.} 2024, \mnras, 533,
  3222, \dodoi{10.1093/mnras/stae2037}

\bibitem[{{Eisenstein} {et~al.}(2025){Eisenstein}, {Johnson}, {Robertson},
  {Tacchella}, {Hainline}, {Jakobsen}, {Maiolino}, {Bonaventura}, {Bunker},
  {Cameron}, {Cargile}, {Curtis-Lake}, {Hausen}, {Pusk{\'a}s}, {Rieke}, {Sun},
  {Willmer}, {Willott}, {Alberts}, {Arribas}, {Baker}, {Baum}, {Bhatawdekar},
  {Carniani}, {Charlot}, {Chen}, {Chevallard}, {Curti}, {DeCoursey},
  {D'Eugenio}, {de Graaff}, {Egami}, {Helton}, {Ji}, {Jones}, {Kumari},
  {L{\"u}tzgendorf}, {Laseter}, {Looser}, {Lyu}, {Maseda}, {Nelson},
  {Parlanti}, {Rauscher}, {Rawle}, {Rieke}, {Rix}, {Rujopakarn}, {Sandles},
  {Saxena}, {Scholtz}, {Sharpe}, {Shivaei}, {Simmonds}, {Smit}, {Topping},
  {{\"U}bler}, {Venturi}, {Williams}, {Witstok}, \& {Woodrum}}]{Eisenstein2025}
{Eisenstein}, D.~J., {Johnson}, B.~D., {Robertson}, B., {et~al.} 2025, \apjs,
  281, 50, \dodoi{10.3847/1538-4365/ae1137}

\bibitem[{{Eisenstein} {et~al.}(2026){Eisenstein}, {Willott}, {Alberts},
  {Arribas}, {Bonaventura}, {Bunker}, {Cameron}, {Carniani}, {Charlot},
  {Curtis-Lake}, {D'Eugenio}, {Ferruit}, {Giardino}, {Hainline}, {Hausen},
  {Jakobsen}, {Johnson}, {Maiolino}, {Rauscher}, {Rieke}, {Rieke}, {Rix},
  {Robertson}, {Stark}, {Tacchella}, {Williams}, {Willmer}, {Baker}, {Baum},
  {Bhatawdekar}, {Boyett}, {Chen}, {Chevallard}, {Circosta}, {Curti},
  {Danhaive}, {DeCoursey}, {Endsley}, {de Graaff}, {Dressler}, {Egami},
  {Helton}, {Hviding}, {Ji}, {Jones}, {Kumari}, {L{\"u}tzgendorf}, {Laseter},
  {Looser}, {Lyu}, {Maseda}, {Nelson}, {Parlanti}, {Perna}, {Pusk{\'a}s},
  {Rawle}, {Rodr{\'\i}guez Del Pino}, {Rujopakarn}, {Sandles}, {Saxena},
  {Scholtz}, {Sharpe}, {Shivaei}, {Silcock}, {Simmonds}, {Skarbinski}, {Smit},
  {Stone}, {Suess}, {Sun}, {Tang}, {Topping}, {{\"U}bler}, {Villanueva},
  {Wallace}, {Whitler}, {Witstok}, \& {Woodrum}}]{Eisenstein2026}
{Eisenstein}, D.~J., {Willott}, C., {Alberts}, S., {et~al.} 2026, \apjs, 283,
  6, \dodoi{10.3847/1538-4365/ae3163}

\bibitem[{{Ellis} {et~al.}(2013){Ellis}, {McLure}, {Dunlop}, {Robertson},
  {Ono}, {Schenker}, {Koekemoer}, {Bowler}, {Ouchi}, {Rogers}, {Curtis-Lake},
  {Schneider}, {Charlot}, {Stark}, {Furlanetto}, \& {Cirasuolo}}]{Ellis2013}
{Ellis}, R.~S., {McLure}, R.~J., {Dunlop}, J.~S., {et~al.} 2013, \apjl, 763,
  L7, \dodoi{10.1088/2041-8205/763/1/L7}

\bibitem[{{Endsley} {et~al.}(2020){Endsley}, {Behroozi}, {Stark}, {Williams},
  {Robertson}, {Rieke}, {Gottl{\"o}ber}, \& {Yepes}}]{Endsley2020}
{Endsley}, R., {Behroozi}, P., {Stark}, D.~P., {et~al.} 2020, \mnras, 493,
  1178, \dodoi{10.1093/mnras/staa324}

\bibitem[{{Fossati} {et~al.}(2017){Fossati}, {Wilman}, {Mendel}, {Saglia},
  {Galametz}, {Beifiori}, {Bender}, {Chan}, {Fabricius}, {Bandara}, {Brammer},
  {Davies}, {F{\"o}rster Schreiber}, {Genzel}, {Hartley}, {Kulkarni}, {Lang},
  {Momcheva}, {Nelson}, {Skelton}, {Tacconi}, {Tadaki}, {{\"U}bler}, {van
  Dokkum}, {Wisnioski}, {Whitaker}, {Wuyts}, \& {Wuyts}}]{fossati17}
{Fossati}, M., {Wilman}, D.~J., {Mendel}, J.~T., {et~al.} 2017, \apj, 835, 153,
  \dodoi{10.3847/1538-4357/835/2/153}

\bibitem[{{Geha} {et~al.}(2012){Geha}, {Blanton}, {Yan}, \& {Tinker}}]{geha12}
{Geha}, M., {Blanton}, M.~R., {Yan}, R., \& {Tinker}, J.~L. 2012, \apj, 757,
  85, \dodoi{10.1088/0004-637X/757/1/85}

\bibitem[{{Graur} {et~al.}(2011){Graur}, {Poznanski}, {Maoz}, {Yasuda},
  {Totani}, {Fukugita}, {Filippenko}, {Foley}, {Silverman}, {Gal-Yam},
  {Horesh}, \& {Jannuzi}}]{Graur2011}
{Graur}, O., {Poznanski}, D., {Maoz}, D., {et~al.} 2011, \mnras, 417, 916,
  \dodoi{10.1111/j.1365-2966.2011.19287.x}

\bibitem[{{Greene} {et~al.}(2024){Greene}, {Labbe}, {Goulding}, {Furtak},
  {Chemerynska}, {Kokorev}, {Dayal}, {Volonteri}, {Williams}, {Wang}, {Setton},
  {Burgasser}, {Bezanson}, {Atek}, {Brammer}, {Cutler}, {Feldmann}, {Fujimoto},
  {Glazebrook}, {de Graaff}, {Khullar}, {Leja}, {Marchesini}, {Maseda},
  {Matthee}, {Miller}, {Naidu}, {Nanayakkara}, {Oesch}, {Pan}, {Papovich},
  {Price}, {van Dokkum}, {Weaver}, {Whitaker}, \& {Zitrin}}]{Greene2024}
{Greene}, J.~E., {Labbe}, I., {Goulding}, A.~D., {et~al.} 2024, \apj, 964, 39,
  \dodoi{10.3847/1538-4357/ad1e5f}

\bibitem[{{Guo} {et~al.}(2017){Guo}, {Bell}, {Lu}, {Koo}, {Faber}, {Koekemoer},
  {Kurczynski}, {Lee}, {Papovich}, {Chen}, {Dekel}, {Ferguson}, {Fontana},
  {Giavalisco}, {Kocevski}, {Nayyeri}, {P{\'e}rez-Gonz{\'a}lez}, {Pforr},
  {Rodr{\'\i}guez-Puebla}, \& {Santini}}]{ycguo17}
{Guo}, Y., {Bell}, E.~F., {Lu}, Y., {et~al.} 2017, \apjl, 841, L22,
  \dodoi{10.3847/2041-8213/aa70e9}

\bibitem[{{Harikane} {et~al.}(2023{\natexlab{a}}){Harikane}, {Ouchi}, {Oguri},
  {Ono}, {Nakajima}, {Isobe}, {Umeda}, {Mawatari}, \& {Zhang}}]{Harikane2023a}
{Harikane}, Y., {Ouchi}, M., {Oguri}, M., {et~al.} 2023{\natexlab{a}}, \apjs,
  265, 5, \dodoi{10.3847/1538-4365/acaaa9}

\bibitem[{{Harikane} {et~al.}(2023{\natexlab{b}}){Harikane}, {Zhang},
  {Nakajima}, {Ouchi}, {Isobe}, {Ono}, {Hatano}, {Xu}, \&
  {Umeda}}]{Harikane2023b}
{Harikane}, Y., {Zhang}, Y., {Nakajima}, K., {et~al.} 2023{\natexlab{b}}, \apj,
  959, 39, \dodoi{10.3847/1538-4357/ad029e}

\bibitem[{{Illingworth} {et~al.}(2013){Illingworth}, {Magee}, {Oesch},
  {Bouwens}, {Labb{\'e}}, {Stiavelli}, {van Dokkum}, {Franx}, {Trenti},
  {Carollo}, \& {Gonzalez}}]{Illingworth2013}
{Illingworth}, G.~D., {Magee}, D., {Oesch}, P.~A., {et~al.} 2013, \apjs, 209,
  6, \dodoi{10.1088/0067-0049/209/1/6}

\bibitem[{{Inayoshi} \& {Maiolino}(2025)}]{Inayoshi2025}
{Inayoshi}, K., \& {Maiolino}, R. 2025, \apjl, 980, L27,
  \dodoi{10.3847/2041-8213/adaebd}

\bibitem[{{Jespersen} {et~al.}(2025){Jespersen}, {Steinhardt}, {Somerville}, \&
  {Lovell}}]{Jespersen2025}
{Jespersen}, C.~K., {Steinhardt}, C.~L., {Somerville}, R.~S., \& {Lovell},
  C.~C. 2025, \apj, 982, 23, \dodoi{10.3847/1538-4357/adb422}

\bibitem[{{Kaib}(2022)}]{Kaib2022}
{Kaib}, N.~A. 2022, Science Advances, 8, eabm9130,
  \dodoi{10.1126/sciadv.abm9130}

\bibitem[{{Kaib} \& {Quinn}(2009)}]{KaibQuinn2009}
{Kaib}, N.~A., \& {Quinn}, T. 2009, Science, 325, 1234,
  \dodoi{10.1126/science.1172676}

\bibitem[{{Khederlarian} {et~al.}(2026){Khederlarian}, {Andrews}, {Newman},
  {Zhang}, \& {Dey}}]{romanphotoz}
{Khederlarian}, A., {Andrews}, B.~H., {Newman}, J.~A., {Zhang}, T., \& {Dey},
  B. 2026, arXiv e-prints, arXiv:2602.10207, \dodoi{10.48550/arXiv.2602.10207}

\bibitem[{{Kocevski} {et~al.}(2023){Kocevski}, {Onoue}, {Inayoshi}, {Trump},
  {Arrabal Haro}, {Grazian}, {Dickinson}, {Finkelstein}, {Kartaltepe},
  {Hirschmann}, {Aird}, {Holwerda}, {Fujimoto}, {Juneau}, {Amor{\'\i}n},
  {Backhaus}, {Bagley}, {Barro}, {Bell}, {Bisigello}, {Calabr{\`o}}, {Cleri},
  {Cooper}, {Ding}, {Grogin}, {Ho}, {Hutchison}, {Inoue}, {Jiang}, {Jones},
  {Koekemoer}, {Li}, {Li}, {McGrath}, {Molina}, {Papovich},
  {P{\'e}rez-Gonz{\'a}lez}, {Pirzkal}, {Wilkins}, {Yang}, \&
  {Yung}}]{Kocevski2023}
{Kocevski}, D.~D., {Onoue}, M., {Inayoshi}, K., {et~al.} 2023, \apjl, 954, L4,
  \dodoi{10.3847/2041-8213/ace5a0}

\bibitem[{{Labb{\'e}} {et~al.}(2023){Labb{\'e}}, {van Dokkum}, {Nelson},
  {Bezanson}, {Suess}, {Leja}, {Brammer}, {Whitaker}, {Mathews}, {Stefanon}, \&
  {Wang}}]{Labbe2023}
{Labb{\'e}}, I., {van Dokkum}, P., {Nelson}, E., {et~al.} 2023, \nat, 616, 266,
  \dodoi{10.1038/s41586-023-05786-2}

\bibitem[{{Lambrides} {et~al.}(2024){Lambrides}, {Garofali}, {Larson}, {Ptak},
  {Chiaberge}, {Long}, {Hutchison}, {Norman}, {McKinney}, {Akins}, {Berg},
  {Chisholm}, {Civano}, {Cloonan}, {Endsley}, {Faisst}, {Gilli}, {Gillman},
  {Hirschmann}, {Kartaltepe}, {Kocevski}, {Kokorev}, {Pacucci}, {Richardson},
  {Stiavelli}, \& {Whalen}}]{lambrides25}
{Lambrides}, E., {Garofali}, K., {Larson}, R., {et~al.} 2024, arXiv e-prints,
  arXiv:2409.13047, \dodoi{10.48550/arXiv.2409.13047}

\bibitem[{{Landy} \& {Szalay}(1993)}]{LandySzalay1993}
{Landy}, S.~D., \& {Szalay}, A.~S. 1993, \apj, 412, 64, \dodoi{10.1086/172900}

\bibitem[{{Lee} {et~al.}(2009){Lee}, {Giavalisco}, {Conroy}, {Wechsler},
  {Ferguson}, {Somerville}, {Dickinson}, \& {Urry}}]{lee09}
{Lee}, K.-S., {Giavalisco}, M., {Conroy}, C., {et~al.} 2009, \apj, 695, 368,
  \dodoi{10.1088/0004-637X/695/1/368}

\bibitem[{{Liu} {et~al.}(2025){Liu}, {Jiang}, {Quataert}, {Greene}, \&
  {Ma}}]{LiuH2025}
{Liu}, H., {Jiang}, Y.-F., {Quataert}, E., {Greene}, J.~E., \& {Ma}, Y. 2025,
  \apj, 994, 113, \dodoi{10.3847/1538-4357/ae0c19}

\bibitem[{{Lyu} {et~al.}(2024){Lyu}, {Alberts}, {Rieke}, {Shivaei},
  {P{\'e}rez-Gonz{\'a}lez}, {Sun}, {Hainline}, {Baum}, {Bonaventura}, {Bunker},
  {Egami}, {Eisenstein}, {Florian}, {Ji}, {Johnson}, {Morrison}, {Rieke},
  {Robertson}, {Rujopakarn}, {Tacchella}, {Scholtz}, \& {Willmer}}]{Lyu2024}
{Lyu}, J., {Alberts}, S., {Rieke}, G.~H., {et~al.} 2024, \apj, 966, 229,
  \dodoi{10.3847/1538-4357/ad3643}

\bibitem[{{Ma} {et~al.}(2025){Ma}, {Greene}, {Setton}, {Goulding},
  {Annunziatella}, {Fan}, {Kokorev}, {Labbe}, {Li}, {Lin}, {Marchesini},
  {Matthee}, {Robbins}, {Sajina}, {Sawicki}, \& {Telford}}]{Ma2025}
{Ma}, Y., {Greene}, J.~E., {Setton}, D.~J., {et~al.} 2025, arXiv e-prints,
  arXiv:2504.08032, \dodoi{10.48550/arXiv.2504.08032}

\bibitem[{{Matthee} {et~al.}(2024){Matthee}, {Naidu}, {Brammer}, {Chisholm},
  {Eilers}, {Goulding}, {Greene}, {Kashino}, {Labbe}, {Lilly}, {Mackenzie},
  {Oesch}, {Weibel}, {Wuyts}, {Xiao}, {Bordoloi}, {Bouwens}, {van Dokkum},
  {Illingworth}, {Kramarenko}, {Maseda}, {Mason}, {Meyer}, {Nelson}, {Reddy},
  {Shivaei}, {Simcoe}, \& {Yue}}]{Matthee2024}
{Matthee}, J., {Naidu}, R.~P., {Brammer}, G., {et~al.} 2024, \apj, 963, 129,
  \dodoi{10.3847/1538-4357/ad2345}

\bibitem[{{McLeod} {et~al.}(2024){McLeod}, {Donnan}, {McLure}, {Dunlop},
  {Magee}, {Begley}, {Carnall}, {Cullen}, {Ellis}, {Hamadouche}, \&
  {Stanton}}]{McLeod2024}
{McLeod}, D.~J., {Donnan}, C.~T., {McLure}, R.~J., {et~al.} 2024, \mnras, 527,
  5004, \dodoi{10.1093/mnras/stad3471}

\bibitem[{{Mu{\~n}oz} {et~al.}(2023){Mu{\~n}oz}, {Mirocha}, {Furlanetto}, \&
  {Sabti}}]{Munoz2023}
{Mu{\~n}oz}, J.~B., {Mirocha}, J., {Furlanetto}, S., \& {Sabti}, N. 2023,
  \mnras, 526, L47, \dodoi{10.1093/mnrasl/slad115}

\bibitem[{{Muzzin} {et~al.}(2013){Muzzin}, {Marchesini}, {Stefanon}, {Franx},
  {McCracken}, {Milvang-Jensen}, {Dunlop}, {Fynbo}, {Brammer}, {Labb{\'e}}, \&
  {van Dokkum}}]{muzzin2013}
{Muzzin}, A., {Marchesini}, D., {Stefanon}, M., {et~al.} 2013, \apj, 777, 18,
  \dodoi{10.1088/0004-637X/777/1/18}

\bibitem[{{Naidu} {et~al.}(2025){Naidu}, {Matthee}, {Katz}, {de Graaff},
  {Oesch}, {Smith}, {Greene}, {Brammer}, {Weibel}, {Hviding}, {Chisholm},
  {Labb\textbackslash'e}, {Simcoe}, {Witten}, {Atek}, {Baggen}, {Belli},
  {Bezanson}, {Boogaard}, {Bose}, {Covelo-Paz}, {Dayal}, {Fudamoto}, {Furtak},
  {Giovinazzo}, {Goulding}, {Gronke}, {Heintz}, {Hirschmann}, {Illingworth},
  {Inoue}, {Johnson}, {Leja}, {Leonova}, {McConachie}, {Maseda}, {Natarajan},
  {Nelson}, {Setton}, {Shivaei}, {Sobral}, {Stefanon}, {Tacchella}, {Toft},
  {Torralba}, {van Dokkum}, {van der Wel}, {Volonteri}, {Walter}, {Wang}, \&
  {Watson}}]{Naidu2025}
{Naidu}, R.~P., {Matthee}, J., {Katz}, H., {et~al.} 2025, arXiv e-prints,
  arXiv:2503.16596, \dodoi{10.48550/arXiv.2503.16596}

\bibitem[{{Naidu} {et~al.}(2026){Naidu}, {Oesch}, {Brammer}, {Weibel}, {Li},
  {Matthee}, {Chisolm}, {Pollock}, {Heintz}, {Johnson}, {Shen}, {Hviding},
  {Leja}, {Tacchella}, {Ganguly}, {Witten}, {Atek}, {Belli}, {Bose}, {Bouwens},
  {Dayal}, {Decarli}, {de Graaff}, {Fudamoto}, {Giovinazzo}, {Greene},
  {Illingworth}, {Inoue}, {Kane}, {Labbe}, {Leonova}, {Marques-Chaves},
  {Meyer}, {Nelson}, {Roberts-Borsani}, {Schaerer}, {Simcoe}, {Stefanon},
  {Sugahara}, {Toft}, {van der Wel}, {van Dokkum}, {Walter}, {Watson},
  {Weaver}, \& {Whitaker}}]{Naidu2026}
{Naidu}, R.~P., {Oesch}, P.~A., {Brammer}, G., {et~al.} 2026, The Open Journal
  of Astrophysics, 9, 56033, \dodoi{10.33232/001c.156033}

\bibitem[{{Oesch} {et~al.}(2018){Oesch}, {Bouwens}, {Illingworth}, {Labb{\'e}},
  \& {Stefanon}}]{Oesch2018}
{Oesch}, P.~A., {Bouwens}, R.~J., {Illingworth}, G.~D., {Labb{\'e}}, I., \&
  {Stefanon}, M. 2018, \apj, 855, 105, \dodoi{10.3847/1538-4357/aab03f}

\bibitem[{{Oesch} {et~al.}(2010){Oesch}, {Bouwens}, {Carollo}, {Illingworth},
  {Trenti}, {Stiavelli}, {Magee}, {Labb{\'e}}, \& {Franx}}]{Oesch2010}
{Oesch}, P.~A., {Bouwens}, R.~J., {Carollo}, C.~M., {et~al.} 2010, \apjl, 709,
  L21, \dodoi{10.1088/2041-8205/709/1/L21}

\bibitem[{{Paquereau} {et~al.}(2025){Paquereau}, {Laigle}, {McCracken},
  {Shuntov}, {Ilbert}, {Akins}, {Allen}, {Arango-Togo}, {Berman},
  {B{\'e}thermin}, {Casey}, {McCleary}, {Dubois}, {Drakos}, {Faisst}, {Franco},
  {Harish}, {Jespersen}, {Kartaltepe}, {Koekemoer}, {Kokorev}, {Lambrides},
  {Larson}, {Liu}, {Le Borgne}, {Lewis}, {McKinney}, {Mercier}, {Rhodes},
  {Robertson}, {Toft}, {Trebitsch}, {Tresse}, \& {Weaver}}]{Paquereau2025}
{Paquereau}, L., {Laigle}, C., {McCracken}, H.~J., {et~al.} 2025, \aap, 702,
  A163, \dodoi{10.1051/0004-6361/202553828}

\bibitem[{{Poggianti} {et~al.}(2010){Poggianti}, {De Lucia}, {Varela},
  {Aragon-Salamanca}, {Finn}, {Desai}, {von der Linden}, \&
  {White}}]{poggianti2010}
{Poggianti}, B.~M., {De Lucia}, G., {Varela}, J., {et~al.} 2010, \mnras, 405,
  995, \dodoi{10.1111/j.1365-2966.2010.16546.x}

\bibitem[{{Pozzetti} {et~al.}(2010){Pozzetti}, {Bolzonella}, {Zucca},
  {Zamorani}, {Lilly}, {Renzini}, {Moresco}, {Mignoli}, {Cassata}, {Tasca},
  {Lamareille}, {Maier}, {Meneux}, {Halliday}, {Oesch}, {Vergani}, {Caputi},
  {Kova{\v{c}}}, {Cimatti}, {Cucciati}, {Iovino}, {Peng}, {Carollo}, {Contini},
  {Kneib}, {Le F{\'e}vre}, {Mainieri}, {Scodeggio}, {Bardelli}, {Bongiorno},
  {Coppa}, {de la Torre}, {de Ravel}, {Franzetti}, {Garilli}, {Kampczyk},
  {Knobel}, {Le Borgne}, {Le Brun}, {Pell{\`o}}, {Perez Montero},
  {Ricciardelli}, {Silverman}, {Tanaka}, {Tresse}, {Abbas}, {Bottini}, {Cappi},
  {Guzzo}, {Koekemoer}, {Leauthaud}, {Maccagni}, {Marinoni}, {McCracken},
  {Memeo}, {Porciani}, {Scaramella}, {Scarlata}, \& {Scoville}}]{Pozzetti2010}
{Pozzetti}, L., {Bolzonella}, M., {Zucca}, E., {et~al.} 2010, \aap, 523, A13,
  \dodoi{10.1051/0004-6361/200913020}

\bibitem[{{Prajs} {et~al.}(2017){Prajs}, {Sullivan}, {Smith}, {Levan},
  {Karpenka}, {Edwards}, {Walker}, {Wolf}, {Balland}, {Carlberg}, {Howell},
  {Lidman}, {Pain}, {Pritchet}, \& {Ruhlmann-Kleider}}]{Prajs2017}
{Prajs}, S., {Sullivan}, M., {Smith}, M., {et~al.} 2017, \mnras, 464, 3568,
  \dodoi{10.1093/mnras/stw1942}

\bibitem[{{Riess} \& {Livio}(2006)}]{RiessLivio2006}
{Riess}, A.~G., \& {Livio}, M. 2006, \apj, 648, 884, \dodoi{10.1086/504791}

\bibitem[{{Ryan} {et~al.}(2017){Ryan}, {Thorman}, {Schmidt}, {Cohen}, {Hathi},
  {Holwerda}, {Lunine}, {Pirzkal}, {Windhorst}, \& {Young}}]{Ryan2017}
{Ryan}, Jr., R.~E., {Thorman}, P.~A., {Schmidt}, S.~J., {et~al.} 2017, \apj,
  847, 53, \dodoi{10.3847/1538-4357/aa85ea}

\bibitem[{{Scholtz} {et~al.}(2024){Scholtz}, {Witten}, {Laporte}, {{\"U}bler},
  {Perna}, {Maiolino}, {Arribas}, {Baker}, {Bennett}, {D'Eugenio}, {Simmonds},
  {Tacchella}, {Witstok}, {Bunker}, {Carniani}, {Charlot}, {Cresci},
  {Curtis-Lake}, {Eisenstein}, {Kumari}, {Robertson}, {Rodr{\'\i}guez Del
  Pino}, {Smit}, {Venturi}, {Williams}, \& {Willmer}}]{Scholtz2024}
{Scholtz}, J., {Witten}, C., {Laporte}, N., {et~al.} 2024, \aap, 687, A283,
  \dodoi{10.1051/0004-6361/202347187}

\bibitem[{{Scholtz} {et~al.}(2025){Scholtz}, {Maiolino}, {D'Eugenio},
  {Curtis-Lake}, {Carniani}, {Charlot}, {Curti}, {Silcock}, {Arribas}, {Baker},
  {Bhatawdekar}, {Boyett}, {Bunker}, {Chevallard}, {Circosta}, {Eisenstein},
  {Hainline}, {Hausen}, {Ji}, {Ji}, {Johnson}, {Kumari}, {Looser}, {Lyu},
  {Maseda}, {Parlanti}, {Perna}, {Rieke}, {Robertson}, {Del Pino}, {Sun},
  {Tacchella}, {{\"U}bler}, {Venturi}, {Williams}, {Willmer}, {Willott}, \&
  {Witstok}}]{Scholtz2025}
{Scholtz}, J., {Maiolino}, R., {D'Eugenio}, F., {et~al.} 2025, \aap, 697, A175,
  \dodoi{10.1051/0004-6361/202348804}

\bibitem[{{Secunda} {et~al.}(2026){Secunda}, {Somerville}, {Jiang}, {Greene},
  {Furtak}, \& {Zitrin}}]{Secunda2026}
{Secunda}, A., {Somerville}, R.~S., {Jiang}, Y.-F., {et~al.} 2026, \apj, 996,
  6, \dodoi{10.3847/1538-4357/ae1f08}

\bibitem[{{Shen} {et~al.}(2024){Shen}, {Zhuang}, {Li}, {Burgasser}, {Fan},
  {Greene}, {Narayan}, {Shapley}, {Sun}, {Wang}, \& {Yang}}]{Shen2024nexus}
{Shen}, Y., {Zhuang}, M.-Y., {Li}, J., {et~al.} 2024, arXiv e-prints,
  arXiv:2408.12713, \dodoi{10.48550/arXiv.2408.12713}

\bibitem[{{Smotherman} {et~al.}(2021){Smotherman}, {Connolly}, {Kalmbach},
  {Portillo}, {Bektesevic}, {Eggl}, {Juric}, {Moeyens}, \&
  {Whidden}}]{Smotherman2021}
{Smotherman}, H., {Connolly}, A.~J., {Kalmbach}, J.~B., {et~al.} 2021, \aj,
  162, 245, \dodoi{10.3847/1538-3881/ac22ff}

\bibitem[{{Somerville} {et~al.}(2025){Somerville}, {Yung}, {Lancaster},
  {Menon}, {Sommovigo}, \& {Finkelstein}}]{Somerville2025}
{Somerville}, R.~S., {Yung}, L.~Y.~A., {Lancaster}, L., {et~al.} 2025, \mnras,
  544, 3774, \dodoi{10.1093/mnras/staf1824}

\bibitem[{{Stone} {et~al.}(2025){Stone}, {Shen}, {Zhuang}, {Hu}, {Pierel},
  {Li}, {Burgasser}, {Greene}, {Pan}, {Shapley}, {Sun}, {Venkatraman}, \&
  {Wang}}]{Stone2025}
{Stone}, Z., {Shen}, Y., {Zhuang}, M.-Y., {et~al.} 2025, arXiv e-prints,
  arXiv:2509.19585, \dodoi{10.48550/arXiv.2509.19585}

\bibitem[{{Sun} {et~al.}(2025){Sun}, {Mu{\~n}oz}, {Mirocha}, \&
  {Faucher-Gigu{\`e}re}}]{Sun2025}
{Sun}, G., {Mu{\~n}oz}, J.~B., {Mirocha}, J., \& {Faucher-Gigu{\`e}re}, C.-A.
  2025, \jcap, 2025, 034, \dodoi{10.1088/1475-7516/2025/04/034}

\bibitem[{{Tacchella} {et~al.}(2023){Tacchella}, {Eisenstein}, {Hainline},
  {Johnson}, {Baker}, {Helton}, {Robertson}, {Suess}, {Chen}, {Nelson},
  {Pusk{\'a}s}, {Sun}, {Alberts}, {Egami}, {Hausen}, {Rieke}, {Rieke},
  {Shivaei}, {Williams}, {Willmer}, {Bunker}, {Cameron}, {Carniani}, {Charlot},
  {Curti}, {Curtis-Lake}, {Looser}, {Maiolino}, {Maseda}, {Rawle}, {Rix},
  {Smit}, {{\"U}bler}, {Willott}, {Witstok}, {Baum}, {Bhatawdekar}, {Boyett},
  {Danhaive}, {de Graaff}, {Endsley}, {Ji}, {Lyu}, {Sandles}, {Saxena},
  {Scholtz}, {Topping}, \& {Whitler}}]{Tacchella2023b}
{Tacchella}, S., {Eisenstein}, D.~J., {Hainline}, K., {et~al.} 2023, \apj, 952,
  74, \dodoi{10.3847/1538-4357/acdbc6}

\bibitem[{{Teplitz} {et~al.}(2013){Teplitz}, {Rafelski}, {Kurczynski}, {Bond},
  {Grogin}, {Koekemoer}, {Atek}, {Brown}, {Coe}, {Colbert}, {Ferguson},
  {Finkelstein}, {Gardner}, {Gawiser}, {Giavalisco}, {Gronwall}, {Hanish},
  {Lee}, {de Mello}, {Ravindranath}, {Ryan}, {Siana}, {Scarlata}, {Soto},
  {Voyer}, \& {Wolfe}}]{Teplitz2013}
{Teplitz}, H.~I., {Rafelski}, M., {Kurczynski}, P., {et~al.} 2013, \aj, 146,
  159, \dodoi{10.1088/0004-6256/146/6/159}

\bibitem[{{Wetzel} {et~al.}(2013){Wetzel}, {Tinker}, {Conroy}, \& {van den
  Bosch}}]{wetzel2013}
{Wetzel}, A.~R., {Tinker}, J.~L., {Conroy}, C., \& {van den Bosch}, F.~C. 2013,
  \mnras, 432, 336, \dodoi{10.1093/mnras/stt469}

\bibitem[{{Whitler} {et~al.}(2025){Whitler}, {Stark}, {Topping}, {Robertson},
  {Rieke}, {Hainline}, {Endsley}, {Chen}, {Baker}, {Bhatawdekar}, {Bunker},
  {Carniani}, {Charlot}, {Chevallard}, {Curtis-Lake}, {Egami}, {Eisenstein},
  {Helton}, {Ji}, {Johnson}, {P{\'e}rez-Gonz{\'a}lez}, {Rinaldi}, {Tacchella},
  {Williams}, {Willmer}, {Willott}, \& {Witstok}}]{Whitler2025}
{Whitler}, L., {Stark}, D.~P., {Topping}, M.~W., {et~al.} 2025, \apj, 992, 63,
  \dodoi{10.3847/1538-4357/adfddc}

\bibitem[{{Willott} {et~al.}(2024){Willott}, {Desprez}, {Asada}, {Sarrouh},
  {Abraham}, {Brada{\v{c}}}, {Brammer}, {Estrada-Carpenter}, {Iyer}, {Martis},
  {Matharu}, {Mowla}, {Muzzin}, {Noirot}, {Sawicki}, {Strait},
  {Rihtar{\v{s}}i{\v{c}}}, \& {Withers}}]{Willott2024}
{Willott}, C.~J., {Desprez}, G., {Asada}, Y., {et~al.} 2024, \apj, 966, 74,
  \dodoi{10.3847/1538-4357/ad35bc}

\bibitem[{{Wiseman} {et~al.}(2021){Wiseman}, {Sullivan}, {Smith}, {Frohmaier},
  {Vincenzi}, {Graur}, {Popovic}, {Armstrong}, {Brout}, {Davis}, {Galbany},
  {Hinton}, {Kelsey}, {Kessler}, {Lidman}, {M{\"o}ller}, {Nichol}, {Rose},
  {Scolnic}, {Toy}, {Zontou}, {Asorey}, {Carollo}, {Glazebrook}, {Lewis},
  {Tucker}, {Abbott}, {Aguena}, {Allam}, {Andrade-Oliveira}, {Annis}, {Bacon},
  {Bertin}, {Brooks}, {Buckley-Geer}, {Burke}, {Carnero Rosell}, {Carrasco
  Kind}, {Carretero}, {Costanzi}, {da Costa}, {Pereira}, {Desai}, {Diehl},
  {Doel}, {Everett}, {Ferrero}, {Flaugher}, {Fosalba}, {Frieman},
  {Garc{\'\i}a-Bellido}, {Gaztanaga}, {Giannantonio}, {Gruen}, {Gruendl},
  {Gschwend}, {Gutierrez}, {Hollowood}, {Honscheid}, {Hoyle}, {James},
  {Krause}, {Kuehn}, {Kuropatkin}, {Maia}, {Marshall}, {Martini}, {Menanteau},
  {Miquel}, {Morgan}, {Ogando}, {Palmese}, {Paz-Chinch{\'o}n}, {Petravick},
  {Pieres}, {Plazas Malag{\'o}n}, {Romer}, {Sanchez}, {Scarpine}, {Schubnell},
  {Serrano}, {Sevilla-Noarbe}, {Soares-Santos}, {Suchyta}, {Swanson}, {Tarle},
  {Thomas}, {To}, {Varga}, {Walker}, \& {DES Collaboration}}]{Wiseman2021}
{Wiseman}, P., {Sullivan}, M., {Smith}, M., {et~al.} 2021, \mnras, 506, 3330,
  \dodoi{10.1093/mnras/stab1943}

\bibitem[{{Wu} {et~al.}(2026){Wu}, {Eisenstein}, {Johnson}, {Hainline},
  {Baker}, {Bunker}, {Cameron}, {Curtis-Lake}, {Danhaive}, {Hausen}, {Helton},
  {Ji}, {Looser}, {Maiolino}, {Mengistu}, {Rinaldi}, {Robertson}, {Sun},
  {Tacchella}, {Trussler}, {Williams}, {Willmer}, \& {Witstok}}]{WuZihao2026}
{Wu}, Z., {Eisenstein}, D.~J., {Johnson}, B.~D., {et~al.} 2026, arXiv e-prints,
  arXiv:2601.15960, \dodoi{10.48550/arXiv.2601.15960}

\end{thebibliography}

\bibliographystyle{aasjournal}

\end{document}